\documentclass{PoS}

\title{Recent STAR results and future prospects of $W^{\pm}$ boson production in polarized 
proton-proton collisions at RHIC}

\ShortTitle{Measurement of $W^{\pm}$ Boson Production at RHIC}

\author{\speaker{Bernd Surrow (for the STAR Collaboration)} \\
        Massachusetts Institute of Technology, 77 Massachusetts Avenue, Cambridge, MA 02139, USA \\
        E-mail: \email{surrow@mit.edu}}

\abstract{
The STAR experiment at the Relativistic Heavy-Ion Collider at Brookhaven National
Laboratory is providing fundamental measurements in high-energy polarized $\vec{p}+\vec{p}$ collisions 
at $\sqrt{s}=200-500\,$GeV to deepen our understanding on the spin structure and
dynamics of the proton. 
This program has completed the first data taking period in 2009 of 
polarized $\vec{p}+\vec{p}$ collisions at $\sqrt{s}=500\,$GeV. This opens a new era in the 
study of the spin-flavor structure of the proton based on the production of $W^{-(+)}$ bosons.
The measurement of the cross section and single-spin asymmetries for midrapidity decay positrons and electrons
from $W^{+}$ and $W^{-}$ boson production in longitudinally polarized $\vec{p}+p$ collisions at $\sqrt{s}=500\,$GeV
is presented.}

\FullConference{35th International Conference on High Energy Physics, ICHEP 2010, July 22-28, 2010, Paris, France}

\begin{document}

\section{Introduction}

Understanding the spin structure of the nucleon remains a fundamental challenge in Quantum Chromodynamics (QCD). 
Experimentally, polarized deep-inelastic scattering 
(pDIS) studies have shown that the quark spins account for only $\approx 33\%$ of the proton spin \cite{Bass:2009dr}. 
Semi-inclusive pDIS measurements \cite{Adeva:1997qz, Airapetian:2004zf, Alekseev:2010ub, Alekseev:2007vi} 
are sensitive to the quark and antiquark spin contributions separated by 
flavor \cite{deFlorian:2008mr, deFlorian:2009vb}. They rely on a quantitative understanding of the fragmentation of 
quarks and antiquarks into observable final-state hadrons. 
While the sum of the contributions from quark and antiquark parton distribution functions (PDFs) 
of the same flavor is well constrained, the uncertainties in the polarized antiquark 
PDFs separated by flavor remain relatively large 
\cite{deFlorian:2008mr, deFlorian:2009vb}.

High energy polarized $\vec{p}+\vec{p}$ collisions at $\sqrt{s}=200-500\,$GeV at RHIC provide a unique way to probe
the proton spin structure and dynamics using hard scattering processes \cite{Bunce:2000uv}. The production of jets and
hadrons is currently the prime focus of the gluon polarization studies. The production of $W^{-(+)}$ bosons at $\sqrt{s}=500\,$GeV
provides an ideal tool to study the spin-flavor structure of the proton. This has been pointed 
out at the very early design stages of the polarized proton collider program at RHIC \cite{Underwood:1991ak}.

The data taking period in 2009 of polarized $\vec{p}+\vec{p}$ collisions at $\sqrt{s}=500\,$GeV opens a new era in the 
study of the spin-flavor structure of the proton based on the production of $W^{-(+)}$ bosons. $W^{-(+)}$ bosons are
produced predominantly through $\bar{u}+d$ $(u+\bar{d})$ collisions and can be detected through
their leptonic decay \cite{Bourrely:1993dd}. 
Quark and antiquark polarized PDFs are probed in calculable leptonic $W$ decays at large
scales set by the mass of the $W$ boson.
The production of $W$ bosons in polarized proton collisions allows for the observation of purely 
weak interactions, giving rise to large, parity-violating, longitudinal single-spin asymmetries.
A theoretical framework has been developed to describe inclusive lepton 
production, $\vec{p}+p \rightarrow W^{\pm}+X \rightarrow l^{\pm}+X$, that can be directly compared with experimental
measurements using constraints on the transverse energy and pseudorapidity of the final-state 
leptons \cite{Nadolsky:2003ga, deFlorian:2010aa}.
This development profits from a rich history of hadroproduction of weak bosons 
at the CERN SPS and the FNAL Tevatron and provides a firm basis to use
$W$ production as a new high-energy probe in polarized $\vec{p}+\vec{p}$ collisions \cite{Kotwal:2008zz}.

\section{Measurement of midrapidity decay $e^{\pm}$ from $W^{\pm}$ boson production}

\begin{figure}[t]
\centerline{\includegraphics[width=130mm]{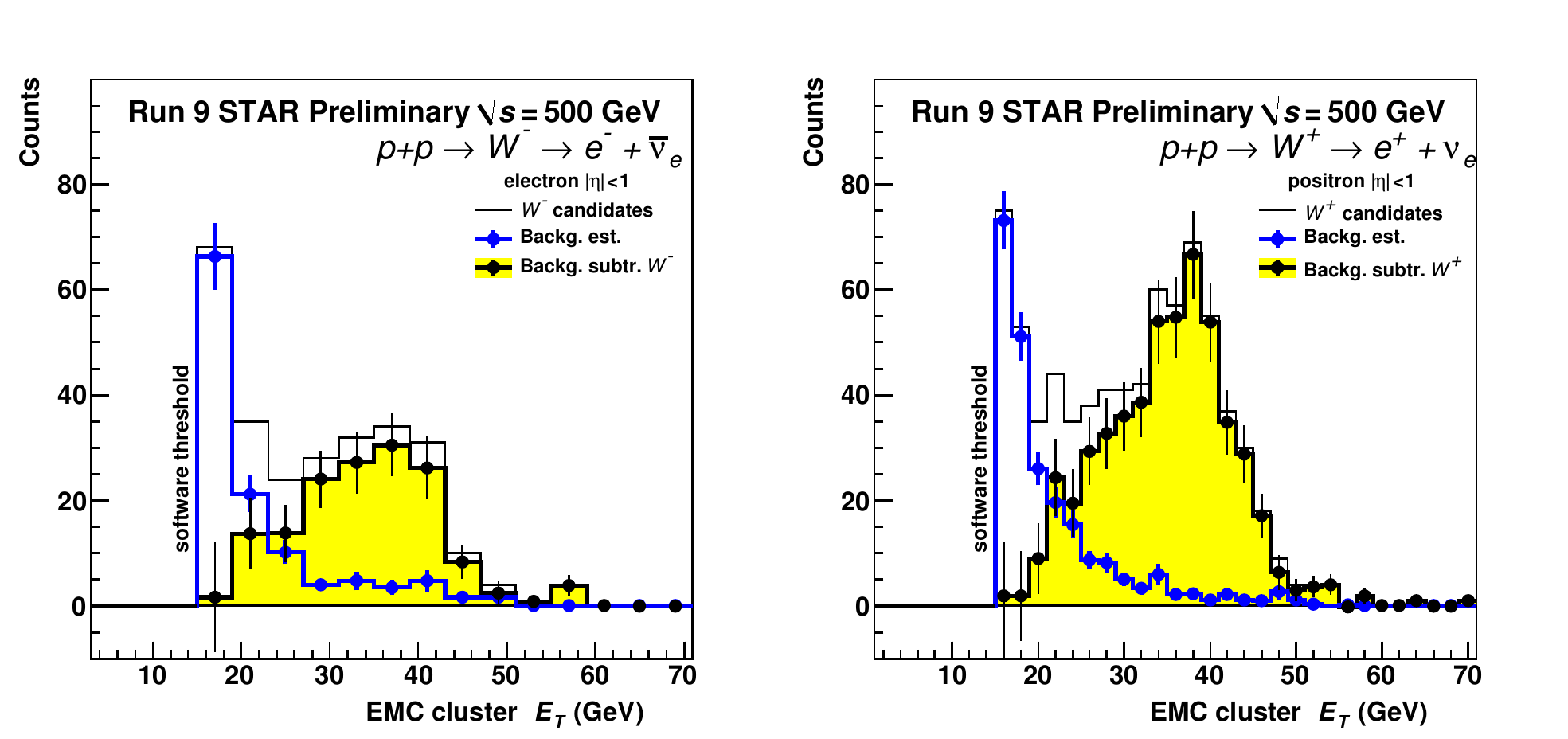}}
\label{fig_j_peak}
\caption{{\it EMC cluster transverse energy distributions for $W^{-}$ (left) and $W^{+}$ (right) events showing the 
candidate histograms in black, the full background estimates in blue and the signal distributions in yellow.}}
\end{figure}

The data used for the first $W$ boson production analysis at STAR were collected in 2009 colliding polarized
proton beams at $250\,$GeV. 
The polarization of each beam was measured 
using Coulomb-Nuclear Interference proton-carbon polarimeters~\cite {Nakagawa:2007zza}, 
which were calibrated using a polarized hydrogen gas-jet target~\cite {Makdisi:2007zz}\@. 
Longitudinal polarization of proton beams in the STAR interaction region (IR) 
was achieved by spin rotator magnets upstream and downstream of the IR that changed the proton spin orientation from its stable vertical direction to 
longitudinal.  
Non-longitudinal beam polarization components were continuously monitored with a local polarimeter system at STAR based
on the Zero-Degree Calorimeters with an upper limit on the relative contribution of $15\%$ for both polarized proton beams \cite{zdc1, zdc3}.   
The longitudinal beam polarizations averaged over all runs 
were $P_1 = 0.38$ and $P_2 = 0.40$ with correlated relative uncertainties 
of 8.3\% and 12.1\%, respectively. 
Their sum $P_1+P_2=0.78$ is used in the analysis and has a relative uncertainty 
of 9.2\%.

Positrons ($e^{+}$) and electrons ($e^{-}$) from $W^{+}$ and $W^{-}$ boson production with $|\eta_{e}|<1$ are 
selected for this analysis.
High-$p_{T}$ $e^{\pm}$ are charge-separated using the STAR TPC.
The BEMC is used to measure the transverse energy $E^{e}_{T}$ of $e^{+}$ and $e^{-}$. 
The suppression of the QCD background is achieved with the TPC, BEMC, and EEMC.

The selection of $W$ candidate events is based on kinematic and topological differences between 
leptonic $W^{\pm}$ decays and QCD background events. Events from $W^\pm$ decays contain a nearly isolated 
$e^{\pm}$ with a neutrino in the opposite direction in azimuth. The neutrino escapes detection leading 
to a large missing energy.
Such events exhibit a large imbalance in the
vector $p_{T}$ sum of all reconstructed final-state objects. In contrast, QCD events, e.g. di-jet events, are characterized 
by a small magnitude of this vector sum imbalance.

Candidate $W$ events were selected online by  
a two-step energy requirement in the BEMC. Electrons or positrons from $W$ production at midrapidity are characterized by
large $E_{T}$ peaked at $\approx M_{W}/2$ (Jacobian peak). At the hardware trigger level, a high tower calorimetric trigger
condition required $E_{T}>7.3\,$GeV in a single BEMC tower. At the software trigger level, a dedicated
trigger algorithm searched for a seed tower of $E_{T}>5\,$GeV and computed all
four possible combinations of $2 \times 2$ tower cluster $E_{T}$ sums above $13\,$GeV. A total of $1.4\times 10^{6}$ events
were recorded for a data sample of approximately  $12\,$pb$^{-1}$. A Vernier scan was used to determine the absolute
luminosity with an accuracy for this preliminary result of currently $23\%$ \cite{APS-Ross}.

An electron candidate is defined to be any TPC track with $p_T>10\,$GeV$/c$ that is associated with a primary vertex 
with $|z|<100\,$cm, where $z$ is the distance along the beam direction.
A $2\times 2$ BEMC tower cluster $E_{T}$ sum, $E_{T}^{e}$,
is required to be larger than $15\,$GeV, referred to as software threshold.
The excess BEMC $E_{T}$ sum in a $4 \times 4$ tower cluster centered around the respective $2\times 2$ tower 
cluster is required to be below $5\%$. In addition, the distance between the $2\times 2$ cluster tower centroid and the TPC track is required to be less 
than $7\,$cm. A near-cone
is formed around the electron candidate direction with a radius in $\eta$-$\phi$ space of $R=0.7$. 
The excess BEMC, EEMC and TPC $E_T$ sum in this cone is required to be less than $12\%$ of the $2\times 2$ cluster $E_T$.
The away-side $E_{T}$ is the EMC plus TPC $E_{T}$ sum over the full $\eta$
range and $\phi \in [\phi_{e} + \phi + 0.7,\, \phi_{e} +\pi-0.7]$. The vector $p_{T}$ sum is defined 
as the sum of the $e^{\pm}$ candidate $p_{T}$ vector and the $p_{T}$ vectors of all the reconstructed 
jets with the thrust axes outside the $R=0.7$ cone around the candidate $e^{\pm}$. Jets were reconstructed using the 
standard mid-point cone algorithm used in previous STAR jet measurements. The final $W$ candidate 
events are selected by requiring the away-side $E_{T}$ to be less than $30\,$GeV and the magnitude of the vector 
$p_{T}$ sum to be larger than 15 GeV.

Figure 1 presents the charge-separated lepton $E^{e}_{T}$ distributions based on the 
selection criteria given above. $W$ candidate events are shown as the black histograms, where the characteristic Jacobian 
peak can be seen at $\approx M_{W}/2$. 

The number of background events was estimated through a combination of a \textsc{pythia} 6.205 \cite{Sjostrand:2000wi} MC simulation
and a data-driven procedure.  
The $e^{+(-)}$ background from $W^{+(-)}$ boson induced $\tau^{+(-)}$ decays was estimated
using a MC simulations
The remaining background is mostly due to QCD dijet events where one of the jets missed the 
STAR acceptance. We have developed a data-driven procedure to evaluate this type of background. 
We excluded the EEMC ($1<\eta<2$) as an active detector in our analysis to estimate the background due to missing 
calorimeter coverage for $-2<\eta<-1$. 
The background contribution due to missing calorimeter coverage along with 
$\tau$ background contributions have been 
subtracted from both $W^{+(-)}$ $E_{T}^{e}$ distributions. 
The remaining background, 
presumably due to missing jets outside the STAR $|\eta|<2$ window, is evaluated 
based on an extrapolation from the region of $E^{e}_{T}<19\,$GeV in both $W^{+(-)}$ $E_{T}^{e}$ 
distributions. The shape is determined from the $E_{T}^{e}$ distribution in events previously rejected
as background.
This shape $E_{T}^{e}$ distribution is normalized to both $W^{+(-)}$ $E_{T}^{e}$ distributions for
$E^{e}_{T}<19\,$GeV.
The total background estimate for $e^{+(-)}$ is shown in Fig. 1 
by the blue line. 

\begin{figure}[t]
\centerline{\includegraphics[width=75mm]{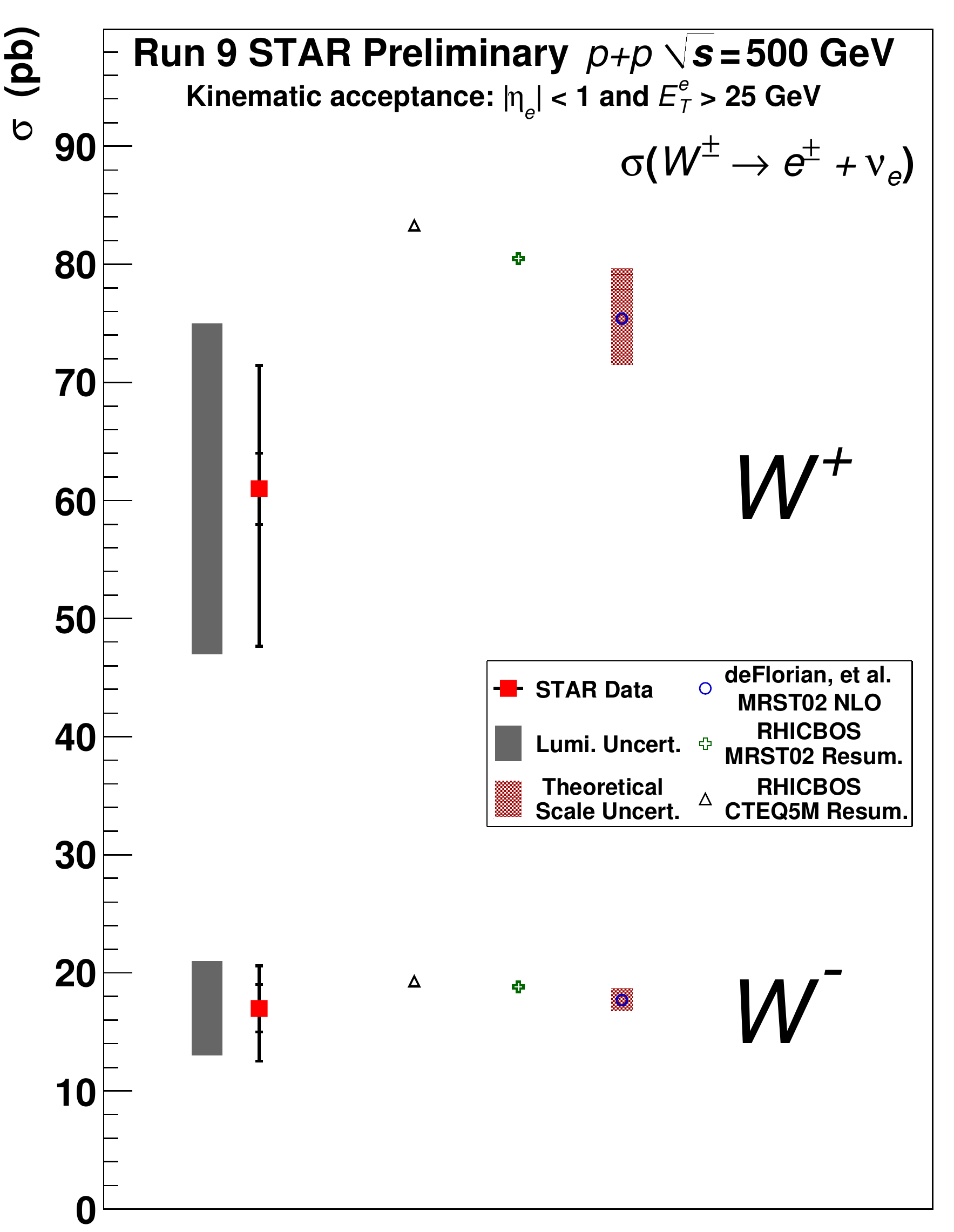}}
\label{fig:xsec}
\caption{{\it Measured cross sections for $W^{\pm}$ production in comparison to predictions based on full resummed 
(RHICBOS)~\cite{Nadolsky:2003ga} and NLO (CHE)~\cite{deFlorian:2005mw} predictions.}}
\end{figure}

\begin{figure}[t]
\centerline{\includegraphics[width=75mm]{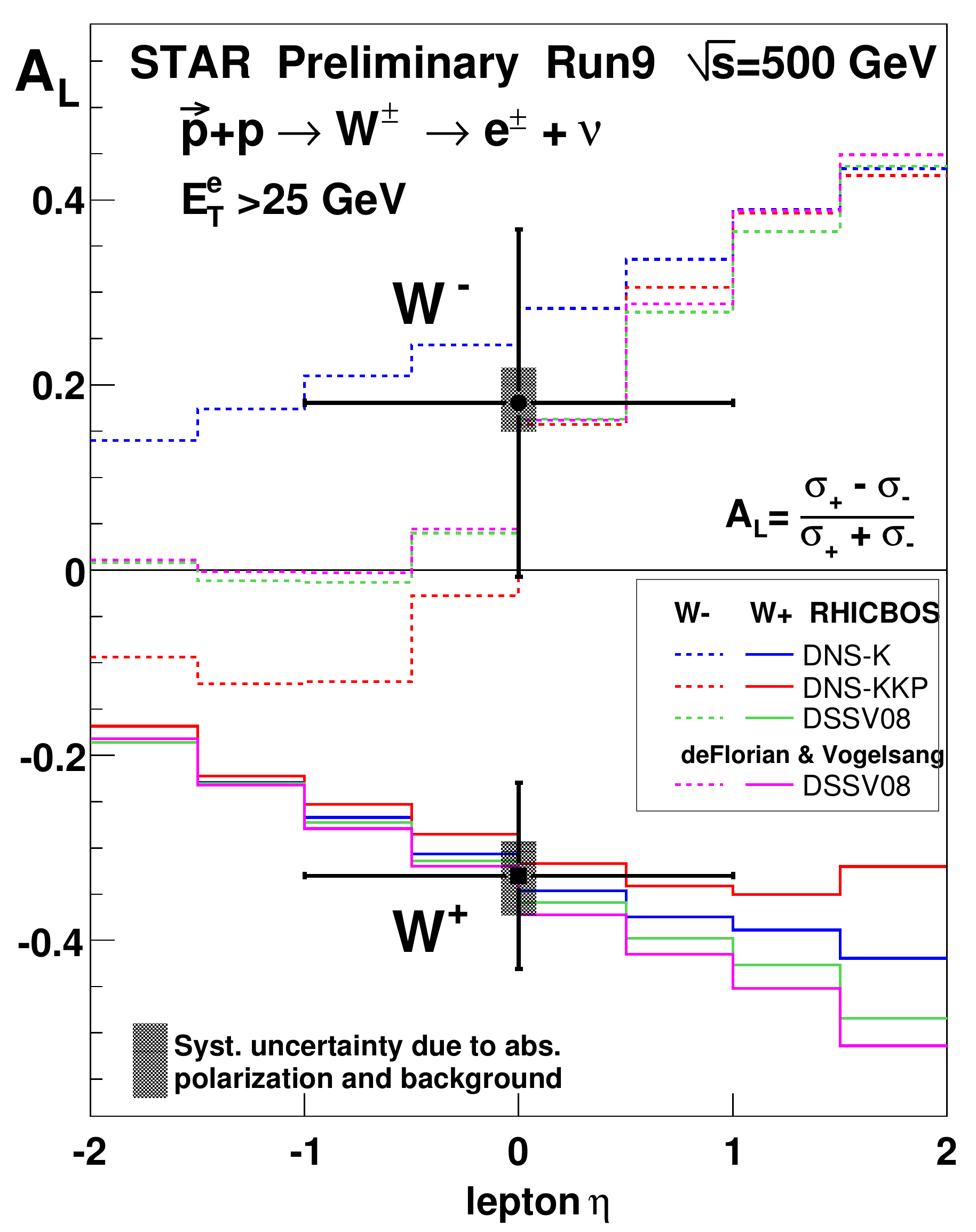}}
\label{fig_wal}
\caption{{\it Longitudinal single-spin asymmetry, $A_{L}$, for $W^{\pm}$ 
events as a function of the leptonic pseudorapidity, $\eta_{e}$, for $25<E^{e}_{T}<50\,$GeV in comparison to theory predictions (See text for details).}}
\end{figure}

\section{Midrapidity $W^{\pm}$ cross section and asymmetry measurements}

Preliminary results for the production cross section of $W^{\pm} \rightarrow e^{\pm}+X$ from candidate events with $|\eta_{e}|<1$ and $E_T^{e}>25$ GeV 
are shown in Fig. 2.  The measured values are $\sigma^{W^{+}}=$ 61 $\pm$ 3 (stat.) $^{+10}_{-13}$ (syst.) 
$\pm$ 14 (lumi.) pb and $\sigma^{W^{-}} =$  17 $\pm$ 2 (stat.) $^{+3}_{-4}$ (syst.) $\pm$ 4 (lumi.) pb.  The statistical and 
systematic uncertainties are shown as the error bars on the red data points.  The systematic uncertainty of the measured luminosity, shown separately 
as the grey bands in Fig. 2, is dominated by the uncertainty in the vernier scan measurement mentioned previously.  The measured 
cross sections are consistent with predictions based on full resummed (RHICBOS)~\cite{Nadolsky:2003ga} and NLO (deFlorian $\&$ Vogelsang)~\cite{deFlorian:2005mw} evaluations, 
which are also shown in Fig. 2. Theoretical scale uncertainties are shown for the NLO predictions as red shaded bands. 

In Figure 3, the measured asymmetries are compared to predictions based on full resummed (RHICBOS) 
\cite{Nadolsky:2003ga} and NLO (deFlorian $\&$ Vogelsang) \cite{deFlorian:2010aa} calculations. The NLO calculations use the DSSV08 polarized PDFs [5], whereas the resummed 
calculations are shown in addition for the older DNS-K and DNS-KKP \cite{deFlorian:2005mw} PDFs.  The NLO and resummed results are in 
good agreement. The range spanned by the DNS-K and DNS-KKP distributions for $\Delta \bar{d}$ and $\Delta \bar{u}$ coincides, 
approximately, with the corresponding DSSV08 uncertainty estimates \cite{deFlorian:2008mr, deFlorian:2009vb}.
The spread of predictions for $A^{W^{+(-)}}_{L}$ is largest at forward (backward) $\eta_{e}$ and is strongly correlated to the 
one found for the $\bar{d}$ ($\bar{u}$) polarized PDFs in the RHIC kinematic region in contrast to the backward (forward)
$\eta_{e}$ region dominated by the behavior of the well-known valence $u$ ($d$) polarized PDFs \cite{deFlorian:2010aa}.
At midrapidity, $W^{+(-)}$ production probes a combination of the polarization of the $u$ and $\bar{d}$ ($d$ and $\bar{u}$) quarks, 
and $A^{W^{+(-)}}_{L}$ is expected to be negative (positive) \cite{deFlorian:2008mr, deFlorian:2009vb}. 
The measured $A^{W^{+}}_{L}$ is indeed negative stressing the direct connection to the $u$ quark polarization. The central value 
of $A^{W^{-}}_{L}$ is positive as expected with a larger statistical uncertainty. Our $A_{L}$ results are consistent with 
predictions using polarized quark and antiquark PDFs constrained by 
inclusive and semi-inclusive pDIS measurements, as expected from the universality of polarized PDFs.

\section{Summary}

In summary, 
we report the first measurement of midrapidity, $|\eta_{e}|< 1$, decay positrons and electrons 
from $W^{+}$ and $W^{-}$ boson production in longitudinally polarized $\vec{p}$+$p$ collisions at $\sqrt{s}=500\,$GeV by the 
STAR experiment at RHIC.
This measurement establishes a new and direct way to explore the spin structure of the 
proton using parity-violating weak interactions in polarized $\vec{p}$+$p$ collisions. 
The measured asymmetries agree well with NLO and resummed calculations using the DSSV08 
polarized PDFs, which are probed at RHIC at much 
larger scales than in previous and ongoing pDIS experiments. 
Future high-statistics measurements at midrapidity together with measurements at forward and 
backward pseudorapidities will focus on constraining the polarization of $\bar{d}$ and $\bar{u}$  quarks.

\bibliographystyle{iopart-num}
\bibliography{surrow}

\end{document}